\documentclass[runningheads,orivec]{llncs}
\usepackage[utf8]{inputenc}
\usepackage[T1]{fontenc}
\usepackage{lmodern}
\usepackage{float}

\usepackage{graphicx}
\usepackage{tikz}
\usepackage{pgfplots}
\pgfplotsset{compat=1.15}
\usepackage{amsmath} 
\usepackage{amsfonts} 
\usepackage{amssymb}
\usepackage[nocompress]{cite}
\usepackage{algorithm}
\usepackage[noend]{algpseudocode}
\usepackage{varwidth}
\usepackage{listings}
\usepackage{xcolor}
\usepackage[hyphens]{url}
\usepackage[hidelinks]{hyperref}
\usepackage[all]{hypcap}
\usepackage{fontawesome}

\usepackage{mathtools}
\usepackage{array}
\usepackage{calc}

\usetikzlibrary{arrows.meta, shapes, fit}

\tikzset{%
	>={Latex[width=2mm,length=2mm]},
	base/.style = {rectangle, rounded corners, draw=black,
		minimum width=2.5cm, minimum height=1cm,
		text centered, font=\sffamily},
	storage/.style = {base, fill=primaryColor},
	symmetric/.style = {base, fill=red!30},
	asymmetric/.style = {base, fill=yellow!30},
	empty/.style = {draw=none},
	pics/vhsplit/.style n args = {5}{
		code = {
			\node[text width=0.5cm] (A) at (#1) {};
			\node[anchor=south west,text width=3cm] (B) at (A.east) {#3};
			\node[anchor=north west,text width=3cm] (C) at (A.east) {#4};
			\node[inner sep=0pt,draw,rounded corners,fit=(A)(B)(C),blend mode=overlay,overlay,fill=#5] (outer) {}; 
			\node[xshift=-1.6cm] (tt) at (outer.center) {#2};  
			\draw (B.north west) -- (C.south west)
			(B.south west) -- (C.north east);    
		}
	},
	pics/vhsplit2/.style n args = {5}{
	code = {
		\node[text width=0.5cm] (A) at (#1) {};
		\node[anchor=south west,text width=3.5cm] (B) at (A.east) {#3};
		\node[anchor=north west,text width=3.5cm] (C) at (A.east) {#4};
		\node[inner sep=0pt,draw,rounded corners,fit=(A)(B)(C),blend mode=overlay,overlay,fill=#5] (outer) {}; 
		\node[xshift=-1.85cm] (tt) at (outer.center) {#2};  
		\draw (B.north west) -- (C.south west)
		(B.south west) -- (C.north east);    
	}
	},
    baseActivity/.style = {rectangle, rounded corners, draw=black, minimum width=4cm, minimum height=.75cm,text centered, font=\sffamily},
	userActivity/.style = {baseActivity, fill=primaryColor},
	serverActivity/.style = {baseActivity, fill=secondaryColor1},
	peerActivity/.style = {baseActivity, fill=secondaryColor2},
}

\colorlet{punct}{red!60!black}
\definecolor{background}{HTML}{F8F8F8}
\definecolor{delim}{RGB}{20,105,176}
\colorlet{numb}{magenta!60!black}
\lstdefinelanguage{JavaScript}{
	morekeywords=[1]{break, continue, delete, else, for, function, if, in,
		new, return, this, typeof, var, void, while, with},
	morekeywords=[2]{false, null, true, boolean, number, undefined,
		Array, Boolean, Date, Math, Number, String, Object},
	morekeywords=[3]{eval, parseInt, parseFloat, escape, unescape},
	sensitive,
	morecomment=[s]{/*}{*/},
	morecomment=[l]//,
	morecomment=[s]{/**}{*/}, % JavaDoc style comments
	morestring=[b]',
	morestring=[b]"
}[keywords, comments, strings]

\lstdefinelanguage{json}{
    basicstyle=\normalfont\ttfamily,
    numbers=left,
    numberstyle=\scriptsize,
    stepnumber=1,
    numbersep=8pt,
    showstringspaces=false,
    breaklines=true,
    frame=lines,
    backgroundcolor=\color{background},
    literate=
     *{0}{{{\color{numb}0}}}{1}
      {1}{{{\color{numb}1}}}{1}
      {2}{{{\color{numb}2}}}{1}
      {3}{{{\color{numb}3}}}{1}
      {4}{{{\color{numb}4}}}{1}
      {5}{{{\color{numb}5}}}{1}
      {6}{{{\color{numb}6}}}{1}
      {7}{{{\color{numb}7}}}{1}
      {8}{{{\color{numb}8}}}{1}
      {9}{{{\color{numb}9}}}{1}
      {:}{{{\color{punct}{:}}}}{1}
      {,}{{{\color{punct}{,}}}}{1}
      {\{}{{{\color{delim}{\{}}}}{1}
      {\}}{{{\color{delim}{\}}}}}{1}
      {[}{{{\color{delim}{[}}}}{1}
      {]}{{{\color{delim}{]}}}}{1},
}

\newcounter{protostepper}

\begin{document}
\title{Partitioned Private User Storages in End-to-End Encrypted Online Social Networks}
\titlerunning{Partitioned Private User Storages}
%\author{anonymized}	
\author{Fabian~Schillinger\orcidID{0000-0001-8771-8290} \and
		Christian~Schindelhauer\orcidID{0000-0002-8320-8581}}

%
%\authorrunning{Author details suppressed}
\authorrunning{F. Schillinger\and C. Schindelhauer}
\institute{Computer Networks and Telematics, \\Department of Computer Science, \\University of Freiburg, Germany\\
	\email{\{schillfa,schindel\}@tf.uni-freiburg.de}}
%\institute{Author details suppressed\\
%\email{Author details suppressed}}
\maketitle              % typeset the header of the contribution

\newcommand{\stealinglogin}{\textit{(S1)}}
\newcommand{\stealingmessages}{\textit{(S2)}}
\newcommand{\stealingkeys}{\textit{(S3)}}

\newcommand{\stealingdata}{\textit{(T1)}}
\newcommand{\modifyingprogram}{\textit{(T2)}}

\newcommand{\alteringkey}{\textit{(I1)}}
\newcommand{\alteringdata}{\textit{(I2)}}
\newcommand{\alteringmetadata}{\textit{(I3)}}

\newcommand{\PBKDF}{\texttt{PBKDF2}}
\newcommand{\RSA}{\texttt{RSA}}
\newcommand{\ECDSA}{\texttt{ECDSA}}
\newcommand{\AES}{\texttt{AES}}
\newcommand{\ECDH}{\texttt{ECDH}}

\newcommand{\storage}{\textit{Storage}}
\newcommand{\storages}{\textit{Storages}}
\newcommand{\storagePart}{\textit{StoragePart}}
\newcommand{\storageParts}{\textit{StorageParts}}

\newcommand{\TSSS}{TS}
\newcommand{\CTSSS}{CTS}

\newcommand{\TSSSKEY}{$K_{\mathit{\TSSS}}$}
\newcommand{\CTSSSKEY}{$K_{\mathit{\CTSSS}}$}
\newcommand{\TSSSSECRET}{$S_{\mathit{\TSSS}}$}
\newcommand{\CTSSSSECRET}{$S_{\mathit{\CTSSS}}$}

%\definecolor{color1a}{RGB}{153,79,136}
%\definecolor{color2a}{RGB}{25,101,176}
%\definecolor{color3a}{RGB}{78,178,101}
%\definecolor{color4a}{RGB}{238,128,38}
%
%\definecolor{color1b}{RGB}{186,122,172}
%\definecolor{color2b}{RGB}{51,139,226}
%\definecolor{color3b}{RGB}{131,201,147}
%\definecolor{color4b}{RGB}{244,170,109}
%
%\definecolor{color1c}{RGB}{103,53,91}
%\definecolor{color2c}{RGB}{15,63,109}
%\definecolor{color3c}{RGB}{54,125,71}
%\definecolor{color4c}{RGB}{185,91,14}

%\definecolor{storagepart1color}{RGB}{95,176,118}
%\definecolor{storagepart2color}{RGB}{227,107,136}
%\definecolor{chatroomcolor}{RGB}{44,111,245}

\definecolor{color1a}{RGB}{159,81,142}
\definecolor{color2a}{RGB}{219,157,111}
\definecolor{color3a}{RGB}{69,136,129}
\definecolor{color4a}{RGB}{184,209,106}

\definecolor{color1b}{RGB}{186,122,172}
\definecolor{color2b}{RGB}{255,205,167}
\definecolor{color3b}{RGB}{104,159,153}
\definecolor{color4b}{RGB}{224,243,159}

\definecolor{color1c}{RGB}{212,167,202}
\definecolor{color2c}{RGB}{255,224,201}
\definecolor{color3c}{RGB}{154,196,191}
\definecolor{color4c}{RGB}{235,248,195}

\definecolor{primaryColor}{RGB}{186,122,172}
\definecolor{secondaryColor1}{RGB}{104,159,153}
\definecolor{secondaryColor1darker}{RGB}{69,136,129}
\definecolor{secondaryColor1lighter}{RGB}{154,196,191}
\definecolor{secondaryColor1darkerdarker}{RGB}{38,110,102}
\definecolor{secondaryColor2}{RGB}{255,233,167}
\definecolor{secondaryColor2darker}{RGB}{219,192,111}
\definecolor{secondaryColor2lighter}{RGB}{255,241,201}

\colorlet{storagepart1color}{secondaryColor2darker}
\colorlet{storagepart1bcolor}{secondaryColor2}
\colorlet{storagepart2color}{secondaryColor2lighter}
\colorlet{chatroomcolor}{secondaryColor1darkerdarker}
\colorlet{chatroomcolor2}{secondaryColor1darker}
\begin{abstract}
In secure Online Social Networks (OSN), often end-to-end encryption approaches are used. This ensures the privacy of communication between the participants. To manage, store, or transfer the cryptographic keys from one device to another one, encrypted private storages can be used. To gain access to such storages, login credentials, only known to the user, are needed. Losing these credentials results in a permanent loss of cryptographic keys and messages because the storage is encrypted. We present a scheme to split encrypted user storages into multiple storages. Each one can be reconstructed with the help of other participants of the OSN. The more of the storages can be reconstructed, the higher the chance of successfully reconstructing the complete private storage is. Therefore, regaining possession of the cryptographic keys used for communication is increased. We achieve high rates of successful reconstructions, even if a large fraction of the distributed shares are not accessible anymore because the shareholders are inactive or malicious.
\end{abstract}

\section{Introduction}
Online Social Networks (OSN) are very popular and the number of monthly users for Facebook, one of the largest OSNs is growing each month and has reached about 2.5 billion monthly users in 2019~\cite{facebookusers}. Participants of OSNs can share content like news articles, personal information, images, and others with friends. Participants further can communicate through OSNs, using integrated online chats or messaging programs. Often the communication is not protected between the users. This allows administrators, attackers, or governmental organizations to read and intercept the communication. Some online chats are protecting the communication of their users by implementing different approaches like end-to-end encryption between the users. WhatsApp is the most popular application, with around 1.6 billion monthly users in 2019~\cite{whatsappusers}. Often, the users have to provide the correct login credentials, like passwords or personal identification numbers to use these applications. If the credentials are lost it can be impossible to retrieve them from the provider, because this would allow undermining the principles of end-to-end encryption. Recovering the credentials can be achieved by implementing secret sharing schemes. 
\subsection*{Our Contribution}
Our contribution consists of an applied secret sharing scheme for end-to-end encrypted online social networks (OSN). The approach allows reconstructing login credentials, even though they are unknown to the server or administrators. To achieve this we rely on encrypted storages, which are partitioned and can be recovered one after another. Each storage contains cryptographic keys for different parts of the OSN. When a storage is recovered the associated part of the OSN can be accessed again. We present algorithms to partition storages, distribute shares for a secret sharing scheme, and to recover the storages. Different approaches and corresponding variables are evaluated in simulations. We suggest optimal values for networks with high fluctuation of active users, which means high numbers of shareholders are inactive and their shares are inaccessible. Further, we discuss the implications on security, given an attack model, where an attacker has full access to the servers of the OSN. The main advantage compared to applied solutions in existing OSNs is that no server or administrator has to know the login credentials, which is essential for end-to-end encryption. Our findings can be transferred to cryptocurrency environments, where similar problems exist.
\subsection*{Organization of the Paper}
The paper is structured as follows: In Section~\ref{Related Work} related work, regarding encrypted OSNs and the reconstruction of login credentials for OSNs is listed. In Section~\ref{Background} the backgrounds on secret sharing schemes and the system model are described. In Section~\ref{Our Approach} our approach is discussed in detail. Further, Section~\ref{Analysis} gives simulation results when applying our scheme and discusses the security implications. Finally, Section~\ref{Conclusion} concludes the work.
\section{Related Work}
\label{Related Work}
Many different solutions to the problem of private communication are known. Often end-to-end-encryption is used. Private keys or passphrases are needed and some of the solutions use private storages for users. Private storages can be used to save data on servers. This data can be retrieved again, but attackers can not access the contents. For some of the solutions, no procedure of how to reconstruct a lost passphrase or private key is shown. Either, because it is not possible or because it lies outside of the scope of the research. 
\subsection{Related Work without Recovery Procedures}\textit{Off-the-record} (OTR)~\cite{OTR1} is a scheme without a special key recovery method. Participants $a$ and $b$ generate multiple private keys $x_{a0}, x_{a1}, \dots$ and $x_{b0}, x_{b1}, \dots$. Using the first keys a common key is generated using a Diffie-Hellmann key exchange. For every new message, the two next keys are used to find a new key.\\ 
\textit{Silent Circle Instant Messaging Protocol} (SCIMP)~\cite{moscaritolo2012silent} is a peer-to-peer approach to send end-to-end encrypted messages between two peers. SCIMP uses \ECDH~to agree on a shared secret. This works as follows: the initiator of a conversation sends a commitment, which is a hash of a newly created \ECDH~public key. The responder answers with a newly generated \ECDH~public key. Hashes of previously shared keys are appended to these two messages, if available. Then, both peers send message authentication codes on a known value. Afterward, the peers can exchange messages which are then encrypted using a \textit{Counter-Mode/CBC-MAC}.\\
\textit{Private Facebook Chat} (PFC)~\cite{robison2012private} is a system to provide secure messaging within the Facebook Chat, by using end-to-end encryption. PFC uses an automated key escrow system, which manages the encryption keys. The keys are distributed by a PFC server, which uses the Facebook authentication mechanism. Users, therefore, do not need an additional password.\\
In~\cite{kikuchi2004secure} a modified Diffie-Hellman (DH) protocol for instant messaging between multiple peers is described. When a peer registers at the server it determines a Diffie-Hellman public key using two secrets from the server and the peer. When a peer initiates a conversation he creates a random number, on which a single message is computed and sent to the server. The server then modifies the message, such that the recipient can perform a DH key exchange using the public key of the initiator. Messages are encrypted using the established key. As long as the server does not know the secrets it can not read the encrypted messages.\\
In~\cite{yang2008design} an instant messaging service based on Elliptic-Curve cryptography is described. The general approach of~\cite{kikuchi2004secure} is used, which includes the registration procedure to generate a public key for a peer, which then is used by other peers to establish a common secret between peers, using \ECDH.\\
The well-known \textit{Signal} protocol~\cite{SIGNAL_doubleratchet_1} uses a shared key which is derived by a key derivation function (KDF) from a KDF chain which itself is derived from another KDF chain. The inputs to this KDF chain are shared secrets found through a modified Diffie-Hellman key exchange. The protocol is used in variations in other applications like \textit{OMEMO}~\cite{OMEMO1}. 
\subsection{Related Work with Recovery Procedure}
Another system using a variation of the Signal protocol is \textit{WhatsApp}~\cite{evans_2014}. It allows to backup messages to a private storage, like Google Drive. When, for example, the original device for using WhatsApp is lost messages can be retrieved again.\\
\textit{SafeSMS}~\cite{1458584} is a tool to exchange end-to-end-encrypted messages through encrypted short message service (SMS) messages. Users can create an encrypted storage on their mobile phone. A possibility to safely store and retrieve the private key could be through storing them on the sim card.\\
\textit{Threema}~\cite{threema} is an end-to-end encrypted online chat. Threema allows users to create a so-called Threema Safe, which stores information of the user like the ID, private key, profile information, and others. The Threema Safe is encrypted using a password chosen by the user and can be stored on the servers of Threema or a private server. Therefore, the Threema Safe can be used to retrieve the private key of a user, but if the password of a user is lost the Threema Safe is useless itself, because it cannot be decrypted without the password.\\ 
\textit{Pretty Good Privacy} (PGP)~\cite{rfc4880} is an approach, where both public-key encryption, as well as, symmetric-key encryption is used for encrypted e-mail communication. The public key is used for key encryption and signature verification and the private key is used for key decryption and signature generation. The message contents are encrypted, using symmetric keys. Some implementations of PGP, like PGP Desktop by Symantec~\cite{symantec} allow users to create five reconstruction questions when installing the PGP Software and generating the key pairs. When a user loses his credentials, the key reconstruction procedure can be used to reconstruct the credentials, if three of the five questions are answered correctly.
\section{Model and Backgrounds}
\label{Background}
For a better understanding, in the following, the term \textit{user} describes a person, which has lost the login credentials for the OSN and wants to recover them. \textit{Recover} and \textit{reconstruct} are used synonymously for the procedure of computing a secret from shares of a secret sharing scheme. A \textit{participant} is a member of a chatroom, which is allowed to read and write messages. The term \textit{peer} describes a participant of a chatroom the user is a member of. I.e. in a chatroom with participants $u$, $p_1$, and $p_2$, where $u$ is the user, both $p_1$ and $p_2$ are peers of $u$. $\{m\}_k$ denotes the ciphertext of message $m$, encrypted with the key $k$. %$[m]$ is the value when a secure hashing algorithm or a key derivation function like \PBKDF~is applied to the message $m$.\\
\subsection{System Model}
\label{System Model}
In this work, we introduce some enhancements for the end-to-end encryption scheme for online chats, as proposed in \cite{schillinger2019end}. A user $u$ owns a keypair $(e_u,d_u)$ for the \RSA~cryptosystem. $e_u$ is distributed via a server and can be used for the encryption of direct messages to $u$. Another key pair $(v_u,s_u)$ owned by $u$ is used for \ECDSA~signatures. Again, $v_u$ is distributed via the server and can be used to verify signatures of $u$. Apart from direct messages between peers, chatrooms can be created. A chatroom can hold arbitrary many participants. Messages sent to a chatroom $c$ are encrypted with a symmetric \AES~key $k_c$. The key $k_c$ is exchanged between the participants using (encrypted) direct messages. I.e. in a chatroom $c$ with peers $p, q, r$ the key $k_c$ is sent to all peers as $\{k_c\}_p, \{k_c\}_q$, and $\{k_c\}_r$. A new key $k'_c$ is exchanged between the participants when the participants of $c$ change. This happens, when a new participant joins or leaves a chatroom. Messages in the chatrooms are signed, as well. Further, a directed acyclic graph is created from the sent and received messages. This allows detecting whether an attacker has deleted messages. All changes in the chatrooms are logged, using \textsc{SystemMessages}. These are special messages in the chatrooms. They are encrypted, signed, and included in the acyclic graph to detect tampering, as well. As long as no attack is detected, these messages are not displayed to the participants. A user $u$ has to store their own keys $d_u$ and $s_u$. Further, $u$ can store the keys $e_i$ and $v_i$ of every other participant $i$ and all keys $k_c$ of the chatrooms $c$. Therefore, $u$ has a private \storage. Using the private login password $P$ of $u$ and a salt value $\mathit{salt}_S$ a symmetric key $P_S = \PBKDF(P+\mathit{salt}_S)$ is derived. Using $P_S$ the \storage~is encrypted and stored on the server. Another salt value $\mathit{salt}_A$ is used to authenticate $u$ against the server with $P_A = \PBKDF(P+\mathit{salt}_A)$. This allows $u$ to authenticate and to decrypt the \storage~using the same password $P$. \PBKDF~is used with \textsc{SHA-256}, therefore calculating $P$ from $P_A$ and $\mathit{salt}_A$ is infeasible and $P_A$ can be stored on the server for the authentication procedure. On the other hand, when $u$ loses his password the server can not help in retrieving $P_S$. All keys in the \storage, then, are inaccessible. %The login procedure is displayed in Algorithm~\ref{alg:login}.
%	\begin{algorithm}[h]
%	\centering
%	\begin{tabular}{@{}p{1.7em}@{}p{0.5\linewidth-0.2em}@{}%
%			p{2.5em}@{}p{0.5\linewidth-4.1em}@{}}
%		%& \textbf{Input:} vectors $c$ and $c''$\\[2ex]
%		%	\hline
%		& & &\\[-1.8ex]
%		& \textbf{User $u$} (with secret password $P$)  & & \textbf{Server}\\[0.5ex]
%		\hline
%		\hline
%		& & &\\[-2ex]
%		\protostep{1}.
%		& wants to login, sends $u$
%		& \sendright \\
%		\protostep{2}.
%		& & \sendleft & sends $\mathit{salt}_A$ and $\mathit{salt}_S$ of $u$\\[0ex]
%		\protostep{3}.
%		& calculates $P_A = \mathit{PBKDF2}(P+\mathit{salt}_A)$  &\\
%		& sends $P_A$ 
%		& \sendright \\
%		\protostep{4}.
%		& &  & compares $P_A$ with stored $P_A'$ for $u$\\[0ex]
%		& \textbf{\boldmath If $P_A = P_A'$:}\\
%		\protostep{5a}.
%		& & \sendleft & sends encrypted $\{\storage\}$~of $u$\\[0ex]
%		\protostep{6a}.
%		& calculates $P_S = \mathit{PBKDF2}(P+\mathit{salt}_S)$ &  \\[0ex]
%		\protostep{7a}.
%		& decrypts $\storage = \mathit{AES}^{-1}(\{\storage\}, P_S)$ &  \\[0ex]
%		& & \centering $\vdots$ &\\[-2ex]
%		& \textbf{\boldmath Else:}\\
%		\protostep{5b}.
%		& &  & denies access, logs failed login attempt\\[0ex]
%		& & &\\[-1ex]
%	\end{tabular}\\[-1ex]
%	\caption{A user $u$ can use a secret password $P$ and two salt values to derive $P_A$ and $P_S$ using \PBKDF. $P_A$ is used to authenticate against the server. With $P_S$ the private \storage~can be decrypted. It is infeasible to calculate $P_S$, even with the knowledge of $P_A$ and the salt values.}
%	\label{alg:login}
%\end{algorithm}
\subsection{Secret Sharing Schemes}
A secret sharing scheme allows distributing a secret, like a password or a private key, to multiple peers, called shareholders. Each shareholder receives a share such that by combining the shares the secret can be recovered. In 1979, Shamir~\cite{shamir1979share} and Blakley~\cite{blakley1979safeguarding} independently came up with the idea of $(t,n)$-threshold secret sharing schemes (\TSSS). In a $(t,n)$-\TSSS~$n$ shares are distributed. Any combination of $t' \geq t$ (different) shares can compute the secret. Knowing $t''<t$ shares gives no advantage in computing the secret compared to guessing it. The scheme of Shamir uses polynomials of degree $t-1$, where $n$ different points are distributed. The secret can be revealed by using polynomial interpolation. $(t-1)$-dimensional hyperplanes are distributed to the shareholders in the scheme of Blakley. By combining $t$ of the $n$ hyperplanes the point of intersection is the secret. Other approaches use Latin squares like the scheme of Cooper~\cite{cooper1994secret}. Here, the secret is a Latin square. Each share is a partial Latin square, such that there exist combinations of shares where the unions constitute critical sets. Each critical set can be used to reconstruct the Latin square. Using the Chinese remainder theorem, other schemes can be constructed. Two approaches are described in~\cite{mignotte1982share} and~\cite{asmuth1983modular}. The basic idea is that the secret $S$ can be computed by using any set of at least $t$ previously chosen coprime integers $m_i$, where the smallest product of any $t$ coprime integers is larger than $S$ and at the same time any set of $t-1$ is smaller. The shares are calculated by $s_i=S \mod m_i$. \\In compartmented threshold secret sharing schemes (\CTSSS) the shareholders are organized in compartments. Each shareholder receives a share to reconstruct the secret of the corresponding compartment. Every secret of a compartment is the share of another secret sharing scheme. Therefore, when enough shares of a compartment are known the compartment's secret can be calculated. By combining the compartments' shares, the secret can be reconstructed. \CTSSS~can be constructed from secret sharing schemes and can contain multiple layers of compartments. Some \CTSSS~are shown in \cite{benaloh1990generalized,ghodosi1998secret,iftene2005compartmented,simmons1988really,lin2009ideal_a}. 
\section{Our Approach}
\label{Our Approach}
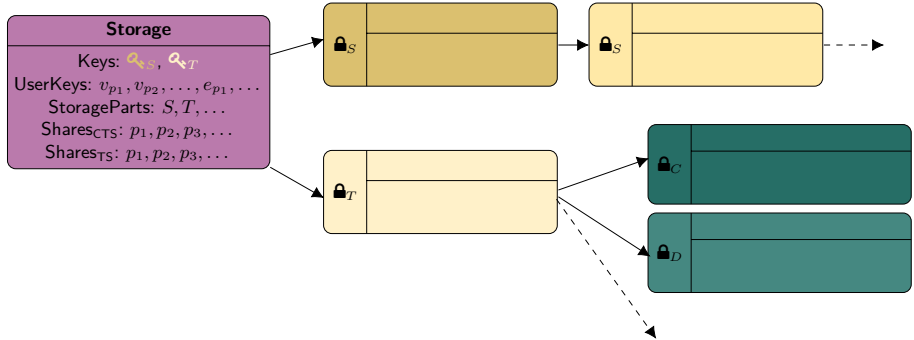
\begin{figure}[htb]
	\centering
	\resizebox{1.0\textwidth}{!}{
		\begin{tikzpicture}[every node/.style={font=\sffamily}, align=center]

		\node (storage)	[storage, rectangle split, rectangle split parts=2] at (-1,0) 
		{
			\textbf{Storage}
			\nodepart{second}Keys: \textcolor{storagepart1color}{\faKey\textsubscript{$S$}}, \textcolor{storagepart2color}{\faKey\textsubscript{$T$}}\\UserKeys: $v_{p_1}, v_{p_2}, \dots, e_{p_1}, \dots$\\StorageParts: $S, T, \dots$\\Shares\textsubscript{\CTSSS}: $p_1, p_2, p_3, \dots$\\Shares\textsubscript{\TSSS}: $p_1, p_2, p_3, \dots$
		};%todo additional share per user

		\node (storagepart1start)	[empty] at (2.5,1) { };
		\path pic (storagepart1) {vhsplit={2.5,1}{\faLock\textsubscript{$S$}}{\textbf{StoragePart $S$}}{Chatrooms: $\dots$\\Keys: $\dots$}{storagepart1color}};
		\node (storagepart1end)	[empty] at (6,0.8) { };
		
		\node (storagepart2start)	[empty] at (7,0.8) { };
		\path pic (storagepart2) {vhsplit={7,1}{\faLock\textsubscript{$S$}}{\textbf{StoragePart $S_1$}}{Chatrooms: $\dots$\\Keys: $\dots$}{storagepart1bcolor}};
		\node (storagepart2end)	[empty] at (10.5,0.8) { };
		
		\node (storagepartempty)	[empty] at (12,0.8) { };
		
		\node (storagepart3start)	[empty] at (2.5,-2.0) { };
		\path pic (storagepart3) {vhsplit={2.5,-1.5}{\faLock\textsubscript{$T$}}{\textbf{StoragePart $T$}}{Chatrooms: $C, D, \dots$\\Keys: \textcolor{chatroomcolor}{\faKey\textsubscript{$C$}} \textcolor{chatroomcolor2}{\faKey\textsubscript{$D$}}, $\dots$}{storagepart2color}};
		\node (storagepart3end)	[empty] at (6,-1.7) { };

		\node (chatroom1start)	[empty] at (8,-1) { };
		\path pic (chatroom1) {vhsplit2={8,-1}{\faLock\textsubscript{$C$}}{\textbf{Chatroom $C$}}{Participants: $p_1, p_2, \dots$\\Messages: $\dots$}{chatroomcolor}};
		
		\node (chatroom2start)	[empty] at (8,-3) { };
		\path pic (chatroom2) {vhsplit2={8,-2.5}{\faLock\textsubscript{$D$}}{\textbf{Chatroom $D$}}{Participants: $p_3, \dots$\\Messages: $\dots$}{chatroomcolor2}};
		
		\node (chatroomempty)	[empty] at (8,-4.5) { };

		%\draw[->]             (start) -- (onCreateBlock);
		%\draw[->]      (activityRuns) -- node[text width=4cm] {Another activity comes in front of the activity} (onPauseBlock);
		%\draw[->] (onPauseBlock.east) -- ++(2.6,0) -- ++(0,2) -- ++(0,2) -- node[xshift=1.2cm,yshift=-1.5cm, text width=2.5cm] {The activity comes to the foreground}(onResumeBlock.east);
		
		\draw[shorten >=0.25cm,shorten <=0.0cm,->] (storage) -- (storagepart1start);
		\draw[shorten >=0.28cm,shorten <=0.0cm,->] (storage) -- (storagepart3start);
		\draw[shorten >=0.25cm,shorten <=0.0cm,->] (storagepart1end) -- (storagepart2start);
		\draw[dashed, shorten >=0.25cm,shorten <=0.0cm,->] (storagepart2end) -- (storagepartempty);
		
		\draw[shorten >=0.25cm,shorten <=0.0cm,->] (storagepart3end) -- (chatroom1start);
		\draw[shorten >=0.25cm,shorten <=0.0cm,->] (storagepart3end) -- (chatroom2start);
		\draw[dashed, shorten >=0.25cm,shorten <=0.0cm,->] (storagepart3end) -- (chatroomempty);
		
		%\draw [dotted] (-6.8,0.5) -- (-6.8,-6.8);
		
		%\draw [dashed, gray] (-10.0,-1) -- (6.0,-1);
		%\draw [dashed, gray, very thick] (-10.5,-4.7) -- (6.0,-4.7);
		\end{tikzpicture}
	}
	\caption{The \storage~holds all keys for the \storageParts. Each \storagePart~contains the keys for chatrooms. Shares for all peers of a user are in the \storage. The shares can recover either a \storagePart~or the \storage. Here, \storagePart~$S$ links to another \storagePart~$S_1$, which holds old keys for the same chatrooms.}
	\label{fig:storagemodel}
\end{figure}

As described in Section~\ref{System Model}, every user $u$ has a private \storage, which is encrypted using a key $P_S$. A private password $P$ is used to derive $P_S$. When $P$ is lost, the stored keys are lost, which results in inaccessible chatrooms. One of the modifications compared to the existing scheme is that $u$ has not just a single \storage, but additional \storageParts. In our scheme, the keys for the chatrooms are moved from the \storage~into the \storageParts. \storageParts~are encrypted, as well. The keys for the \storageParts~are generated randomly by $u$, when they are created. The \storage~holds the keys for the \storageParts. Two secret sharing schemes are used to distribute $P_S$: $u$ generates two random keys \CTSSSKEY~and \TSSSKEY. $u$ then calculates \CTSSSSECRET~$=$~\AES($P_S$, \CTSSSKEY)~$= \{P_S\}_{K_{\mathit{CTS}}}$ and \TSSSSECRET~$=$~\AES($P_S$, \TSSSKEY)~$= \{P_S\}_{K_{\mathit{TS}}}$. A \TSSS~is used to distribute shares of \TSSSSECRET~to all peers the user knows. An additional \CTSSS~is used to distribute shares of \CTSSSSECRET~to all \storageParts. The secret distributed inside a \storagePart~is the key, which is used to encrypt it. I.e. the \storageParts~are compartments for a \CTSSS. This allows reconstructing the \storageParts~one by one, and in parallel to reconstruct the \storage. \CTSSSKEY~and \TSSSKEY~are sent to the server. An overview of a partitioned \storage~is displayed in Figure~\ref{fig:storagemodel}.
%An overview of the scheme is displayed in Figure~\ref{fig:scheme}.
%\begin{figure}[H]
%\centering
%\includegraphics[width=1\linewidth]{scheme_paper}
%\caption[The proposed scheme, to reconstruct private user storages]{The private \storage~of a user contains the keys for \RSA~decryption, \ECDSA~signatures, and the unique \AES~key for every \storagePart. Further, all shares for other participants of the online chat are stored. The \storage~is encrypted using a key $P_S$, which is only known to the user. Each \storagePart~is encrypted using a unique \AES~key, which is stored in the \storage. Therefore, access to the \storage~gives access to all \storageParts~and shares of other users. Each \storagePart~contains the newest \AES~keys for some chatrooms. Older keys are stored in, so called, predecessors. Each participant of a chatroom receives a share for a \TSSS. These shares can be used to reconstruct the according \storagePart. Each \storagePart has the share for a \TSSS, which can be used to reconstruct the \storage.}
%\label{fig:scheme}
%\end{figure}
\subsection{Distributing the Shares}
The key $P_S$ is used to encrypt and decrypt the \storage. $P_S$ is derived from the secret password $P$ of a user $u$. Therefore, $P_S$ has to be distributed using secret sharing schemes. In the \storage~a symmetric key $k_{\mathit{SP}}$ exists for every \storagePart~$\mathit{SP}$. Two keys \CTSSSKEY~and \TSSSKEY~are generated by $u$ randomly. The two keys are sent to the server. In case $u$ has to recover $P_S$ the server helps and provides \CTSSSKEY~and \TSSSKEY. $u$ calculates \CTSSSSECRET~$= $\AES($P_S$, \CTSSSKEY) and \TSSSSECRET~$= $\AES($P_S$, \TSSSKEY). \CTSSSSECRET~and \TSSSSECRET, therefore, are both different ciphertexts of $P_S$. \TSSSSECRET~is distributed as follows: $u$ creates a list $L$ of all peers it knows. By iterating through all \storageParts~and the contained chatrooms this list can be filled. A \TSSS~then is used to create a share $\mathcal{S}_p$ of \TSSSSECRET~for every peer $p \in L$. Then, $\mathcal{S}_p$ is encrypted using the public key $e_p$ of $p$ and signed by $u$. $u$, then, sends the ciphertext and signature to $p$. $p$ then can verify the signature, decrypt $\mathcal{S}_p$, and save it to the own \storage.\\
The other secret \CTSSSSECRET~is distributed by first creating a share $\mathcal{S}_{\mathit{SP}}$ for every \storagePart~$\mathit{SP}$, using a \TSSS. In a \storagePart~$\mathit{SP}$, at first, the symmetric key $k_{\mathit{SP}}$, used to encrypt and decrypt the \storagePart~is encrypted, using the key \CTSSSKEY. Then, a list $L_{\mathit{SP}}$ of all peers of the chatrooms in $\mathit{SP}$ is created. Using another \TSSS, a share $\mathcal{S}_{\mathit{SP},p}$ is calculated for every peer $p\in L_{\mathit{SP}}$. This share, again, is encrypted with $e_p$, signed by $u$, and sent as a direct message to $p$. Therefore, this \TSSS~later allows reconstructing the key for the \storagePart. The distribution of shares is displayed in Algorithm~\ref{algo:constructingSS}.\\
\begin{algorithm}[htb]
	\caption{The algorithm distributes a key to all peers of a user with a secret sharing scheme. For every \storagePart~another secret sharing scheme is used to distribute this key to the users.}
	\label{algo:constructingSS}
	\begin{algorithmic}[1]
		\Procedure{createAndDistributeShares}{\null}
		\State{$S_{\mathit{CTS}}\gets$\textsc{AES}($P_S$, $K_{\mathit{CTS}}$)\Comment{where $K_{\mathit{CTS}}$ is random and $P_S$ is the secret key}}
		\State{$S_{\mathit{TS}}\gets$\textsc{AES}($P_S$, $K_{\mathit{TS}}$)\Comment{where $K_{\mathit{TS}}$ is random}}
		\State{$L \gets \emptyset$}
		\State{$\mathcal{S}\gets$\textsc{SecretSharing}($S_{\mathit{CTS}}$, $\mathit{StorageParts}$)\Comment{create shares $\mathcal{S}$ for \storageParts}}
		\ForAll{\storagePart~$\mathit{SP}$ in \storage}
		\State{$\mathit{SP}.\mathit{Share}\gets \mathcal{S}_{\mathit{SP}}$\Comment{where $\mathcal{S}_{\mathit{SP}}$ is a share for $\mathit{SP}$ from $\mathcal{S}$}}
		\State{$L \gets L \cup L_{\mathit{SP}}$\Comment{where $L_{\mathit{SP}}$ contains all peers from the chatrooms in $\mathit{SP}$}}
		\State{$S\gets$\textsc{AES}($k_{\mathit{SP}}$, $K_{\mathit{TS}}$)}\Comment{where $k_{\mathit{SP}}$ is the symmetric key for $\mathit{SP}$}
		\State{$\mathcal{S}_{k_\mathit{SP}}\gets$\textsc{SecretSharing}($S$, $L_{\mathit{SP}}$)}
		\ForAll{$p \in L_{\mathit{SP}}$}
		\State{\textsc{sendShare}($\mathcal{S}_{\mathit{SP}, p}$, $p$) \Comment{where $\mathcal{S}_{\mathit{SP}, p}$ is the share of $p$ from $\mathcal{S}_{k_\mathit{SP}}$}}
		\EndFor
		\EndFor
		\State{$\mathcal{S}_L\gets$\textsc{SecretSharing}($S_{\mathit{TS}}$, $L$)\Comment{create shares $\mathcal{S}$ for peers}}
		\ForAll{$p \in L$}
		\State{\textsc{sendShare}($\mathcal{S}_{\mathit{p}}$, $p$) \Comment{where $\mathcal{S}_{\mathit{p}}$ is the share of $p$ from $\mathcal{S}_L$}}
		\EndFor
		\EndProcedure 
		\Procedure{sendShare}{$\mathcal{S}$, $p$}
		\State{$c_\mathcal{S}\gets$\textsc{RSA}($\mathcal{S}$, $e_p$)\Comment{where $e_p$ is the public key of $p$}}
		\State{$s\gets$\textsc{Sign}($c_\mathcal{S}$, $s_u$)\Comment{where $s_u$ is the signing key of the user}}
		\State{send ($c_\mathcal{S}$,$s$) to $p$}
		\EndProcedure
	\end{algorithmic}
\end{algorithm}
\subsection{Recovering the Storage}
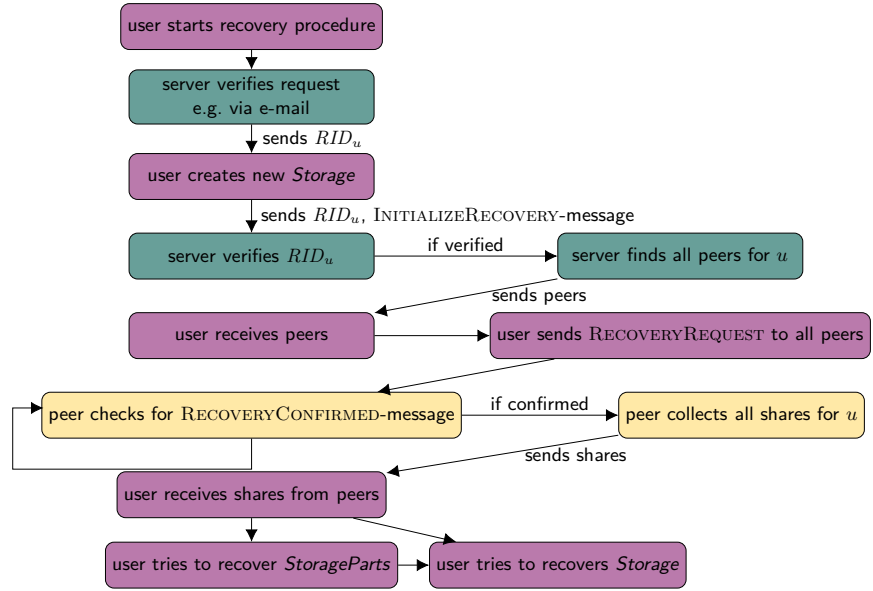
\begin{figure}[htb]
	\centering
	\resizebox{0.95\textwidth}{!}{
		\begin{tikzpicture}[every node/.style={font=\sffamily}, align=center]
		
			\node (a)             [userActivity]              {user starts recovery procedure};
			\node (b)     [serverActivity, below of=a, yshift=-.15cm]          {server verifies request\\e.g. via e-mail};
			\node (c)      [userActivity, below of=b, yshift=-0.3cm]   {user creates new \storage};
			\node (d)     [serverActivity, below of=c, yshift=-0.3cm]   {server verifies $\mathit{RID}_u$};
			\node (e)      [serverActivity, right of=d, xshift=6.0cm]   {server finds all peers for $u$};
			\node (f)      [userActivity, below of=d, yshift=-0.3cm]	{user receives peers};
			\node (g)       [userActivity, right of=f, xshift=6.0cm]	{user sends \textsc{RecoveryRequest} to all peers};
			\node (h)    [peerActivity, below of=f, yshift=-0.3cm] {peer checks for \textsc{RecoveryConfirmed}-message};
			\node (i)    [peerActivity, right of=h, xshift=7.0cm]	{peer collects all shares for $u$};
			\node (j)      [userActivity, below of=h, yshift=-0.3cm]	{user receives shares from peers};
			\node (l) [userActivity, below of=j, yshift=-.15cm]	{user tries to recover \storageParts}; 
			\node (k) [userActivity, right of=l, xshift=4cm]	{user tries to recovers \storage};    
	
			% Specification of lines between nodes specified above
			% with aditional nodes for description 
			\draw[->]             (a) -- (b);
			\draw[->]     (b) -- node[xshift=1.0cm] {sends $\mathit{RID}_u$} (c);
			\draw[->]      (c) -- node[xshift=3.2cm] {sends $\mathit{RID}_u$, \textsc{InitializeRecovery}-message}(d);
			\draw[->]     (d) -- node[yshift=.2cm] {if verified} (e);
			\draw[->]      (e) -- node[xshift=1.2cm] {sends peers} (f);
			\draw[->]      (f) -- (g);
			\draw[->]    (g) -- (h);
			\draw[->]       (h) -- node[yshift=.2cm] {if confirmed} (i);
			\draw[->]    (i) -- node[xshift=1.2cm] {sends shares} (j);
			\draw[->]    (j) -- (k);
			\draw[->]    (j) -- (l);
			\draw[->]    (l) -- (k);
			\draw[->] (h.south) -- ++(0,-0.5) -- ++(-3.9,0) -- ++(0,1.0) -- ++(0.5,0);
		\end{tikzpicture}
	}
	\caption{The user $u$ initializes the recovery procedure. The help of the server and some of the peers is needed to recover the \storage~or multiple \storageParts. The \textsc{RecoveryConfirmed}-message can be created by any peer $p$ of $u$. To create it, both of them have to meet: $u$ presents a signed message containing $\mathit{RID}_u$ to $p$. $p$ in return, verifies it, signs it again, and sends it to the server.}
	\label{fig:recovery}
\end{figure}

When a user $u$ loses his password $P$ he can neither log in to the OSN nor decrypt the \storage. To access the \storage~again the following procedure is used: At first, the user tells the server to allow him to get an onetime login. This should be allowed only if the user can provide evidence, that he is the owner of the account. This can be achieved, for example by sending the user a link via e-mail, which he can click only if he knows the password to the mailbox. There are different ways of achieving this, which depend on the OSN, where the chat is used. The server creates a new ID $\mathit{RID}_u$ of the recovery request and sends it to $u$. Then, $u$ creates a new \storage~and sends an \textsc{InitializeRecovery}-message containing $\mathit{RID}_u$ to the server. The server searches for all peers $P$ of $u$. This can be done because the lists of participants for all chatrooms are stored on the server in plaintext. $P$ is sent to $u$, together with \TSSSKEY~and \CTSSSKEY. Then, $u$ sends a \textsc{RecoveryRequest}-message to every peer $p\in P$, again containing $\mathit{RID}_u$. When $p$ receives such a message it checks whether a \textsc{RecoveryConfirmed}-message for $\mathit{RID}_u$, signed by a participant of the OSN exists. If no such message exists $p$ interrupts the procedure and checks for the presence of the message regularly. If such a message exists, all shares for $u$ are searched in the \storage. The shares are encrypted using the new public key of $u$ and signed by $p$. Then, $p$ sends a \textsc{SystemMessage} to all chatrooms, where $p$ and $u$ are participating in. This message contains the IDs of the \textsc{RecoveryRequest}-message and $u$. The message is encrypted, using the latest \AES~key of the corresponding chatroom. This allows every participant to see, that $u$ wants to recover the key. When $u$ receives a share, it tries to recover, either, the corresponding \storagePart, via the \CTSSS~or the \storage, via the \TSSS. When a secret is calculated $u$ has to decrypt it, using the corresponding key \TSSSKEY~or \CTSSSKEY. When a \storagePart~is recovered $u$ can append it to the new \storage. This allows $u$ to already communicate in some chatrooms. The \storage~can be recovered by $u$, when enough shares of the \TSSS~are combined, or when enough \storageParts~are recovered by $u$. When the \storage~finally is recovered, a \textsc{recoveryFinished}-message is sent to all peers. This is useful for preventing peers that did not send their shares yet, from sending them late. \\The previously mentioned \textsc{RecoveryConfirmed}-message is a message created by any peer $p$ of $u$. The message is created, when $u$ contacts $p$ over a different channel. Such a channel can be a verified e-mail, phone, or a meeting in person. First, $u$ provides a signed message containing $\mathit{RID}_u$ to $p$. This can be achieved by displaying a QR code or presenting a human-readable version of the message, computed in a way like described in~\cite{dechand2016empirical}. $p$ then verifies the signature, signs the message, and sends it as a \textsc{RecoveryConfirmed}-message to the server. In this way, each peer can check whether the recovery request was started by $u$, and at the same time, $p$ guarantees the correctness. In Figure~\ref{fig:recovery} the recovery procedure is shown as a flowchart.

\section{Analysis of the Proposed Scheme}
\label{Analysis}
When the \storage~is split into different parts, on the one hand, the chance of a successful recovery is increased, but on the other hand, some overhead in the size is generated. Further, additional attack vectors may be introduced. This is analyzed in the following.

\subsection{Considerations on the Overhead of Disk Space}
The \storageParts~are linked with their ID, \AES~key, hash value and share in the \storage. When saving the \storage~as a JSON file, this link is around 320~bytes per \storagePart, when encrypting with \AES~this results in around 580~bytes. For each share, another 30~bytes in plaintext, or 90~bytes in ciphertext are stored. Each \storagePart~has around 370~bytes, when ignoring the list of chats and chat keys. Therefore, an encrypted \storagePart~needs about 610~additional bytes. The list of chats and chat keys is needed, no matter, whether they are stored in the \storage~or a \storagePart. An additional value, indicating the last distribution, needs around 180~bytes when encrypted. An average user of an online chat is in contact with around 70 peers in 60 chatrooms, corresponding to \cite{seufert2015analysis,rosenfeld2018study}. Splitting up the \storage~into four \storageParts~results in an overhead of around ${4\cdot (580+610)+70\cdot 90 - 180 = 10.880}$~bytes. When considering, that storing a single key for a chat needs about 380~bytes, when encrypted, this results in about 22.800~bytes for the same average user. The total overhead, when using our approach is below $\frac{1}{2}$ times the size, the stored keys for this user need. When more then one key per chat exists the ratio gets smaller. This can happen if the users of chatrooms change. This overhead is justifiable because the chance to successfully recover the \storage~is increased. 

%Ein AES-Key hat 122 Zeichen stringified
%share ist in der gleichen größenordnung
%hash256 ist 32Zeichen -> 37bytes
%hash512 ist 64Zeichen -> 100bytes
%
%in storage zu sparen: storageParts: [{id: number, key: object, hash: string, share: numer}]
%in shares zu sparen: storagePart: Number
%in storagePart zu sparen: id, share, predeecssorId
%weiterhin nur noch 1x nötig: lastUpdate, threshold, totalkeys
%
%wenn die chats ausgelagert werden:
%in storage: share: number
%in storagepart: share
%nur noch 1x last update
%threshold, totalkeys%
%
%größe der storageparts, abhängig von den keys die gespieichert sind
%vergleich storage komplett - storageparts\\
%storageparts - storageparts aufgeteilt auf mehrere\\
%storageparts - history\\%
%
%wie groß ist sha256 hash?\\
%wie groß ist ein key?\\
%was ist der overhead?\\
%wie groß ist dann was verschlüsseltes?\\
%TODO: immer der gleiche share pro user?
%TODO: immer nur der seed für einen rng? damit spart man sich alle details abzuspeichern
\subsection{Discussion of Possible Attacks}
\label{attackmodel}
A key recovery method can provide additional attack vectors for someone who wants to access the private messages of a user. An attacker in the system can be any person with full access to the servers of the OSN. It can be an administrator, hacker, or a governmental organization agent with full access to all data. Then, the attacker can read, modify, or delete data on the server. Additionally, any participant of the OSN can be an attacker, as well. Attacks on the client machines are not covered by the scheme. In the following, possible attacks are analyzed in detail.\\
\textbf{Compromised Server}
A compromised server can initiate the recovery procedure. All messages from and to the user $u$ can be intercept and read, because new keys are generated when the recovery procedure is started. The server knows all peers of $u$ and can send them the appropriate messages to receive the shares. At this point the attack fails, because no \textsc{RecoveryConfirmed}-message exists. The server cannot generate such a message without the help of another participant, because it has to be signed by $u$ and another participant.\\
\textbf{Colluding Participants}
When participants of the OSN collude they can combine their shares and compute \TSSSSECRET, or \CTSSSSECRET. In principle, this could work, because the list of participants is visible to anybody inside a chatroom. But, not all peers of a user may be reached, because some of the chatrooms may be unknown to the colluding participants. If nevertheless, enough of them work together they still need \TSSSKEY~or \CTSSSKEY~from the server. In practice, this would mean that, depending on the setting, about 75\% of all peers of a user and an administrator or hacker on the server have to work together.\\
\textbf{Stranger Friends}
In OSN users usually have more and more stranger friends, which are peers that are not known in person by the user. These stranger friends may even be malicious. Therefore, collusion is even more likely. But again, at first enough participants have to collude. At second, all non-stranger friends can detect and block the recovery process, because the \textsc{RecoveryConfirmed}-message is incorrect. They can inform the server and the user.\\
\textbf{Colluding Participants and Server}
When an attacker on the server is colluding with multiple participants the chance to successfully reconstruct the \storage~while staying concealed is higher. The server knows all peers of a user, therefore it is easy to contact all peers to gain their shares. If enough of the peers help, the reconstruction can be hidden by the peers and the server, because either the correct messages can be deleted, or fake \textsc{SystemMessages} can be used to replace the critical messages, which are originally used to reveal all malicious recovery procedures. The question is, whether this effort is justified: First, the server gets to know the communication of the colluding peers, when the \storage~is recovered. Second, when enough peers work together with the server it is far simpler to just forward all messages of the user, instead of recovering the \storage.\\
\textbf{Attack of a Third Party}
It is ensured, that no stranger starts the recovery procedure. At first, the attacker has to get access to the mailbox to reset the password for the OSN. Second, the attacker has to fake a \textsc{RecoveryConfirmed}-message. To achieve this, the attacker has to have access to another account, or he has to work together with a peer of the user. On the one hand, the procedure generates \textsc{SystemMessages} in all chatrooms, which make it traceable, who was attacking the user. On the other hand, the chance of a successful attack can further be reduced, if a \textsc{RecoveryConfirmed}-message has to be signed by multiple peers.\\
\textbf{Stealing the User Login Data}
There are two different keys $P_S$ and $P_A$, both derived using \PBKDF~from the password $P$ of the user. $P_A$ is known to the server. Therefore, it is known to an attacker, as well, whereas $P_S$ is secret. An attacker, which initializes the recovery procedure can set a new password $P'$. This password then is used to derive $P'_A$ and $P'_S$. Now, the attacker knows $P'_S$ which does not help in computing $P_S$. Without the recovered \storage~an attacker gained nothing. The recovery procedure itself is secured against an attacker.\\
\textbf{Stealing Messages and Cryptographic Keys}
An attacker has to recover the \storage~of a user to steal messages or cryptographic keys. The recovery procedure is secured, and every tampering, modification, or started recovery procedure is detected through \textsc{SystemMessages} in all chatrooms. \textsc{SystemMessages} are encrypted and signed, using the correct keys. Further, every deletion of a \textsc{SystemMessage} can be detected, because the message history becomes inconsistent.\\
\textbf{Modifying Cryptographic Keys} 
Cryptographic keys can only be modified, when the \storage~is successfully recovered by an attacker. Such an attack can be revealed.\\
\textbf{Altering Communication Data and Metadata}
Recovering the \storage~as an attacker gives no advantage for deleting or modifying messages. This is not possible in the described scheme. Adding messages, on the other hand, is possible, when the attacker knows the keys, just like that, new participants can be added by an attacker to the chatrooms and old participants can be excluded. This is visible to all participants and therefore can be prevented or be undone by honest participants.\\
\textbf{Small Rooms or Few Peers}
When a user is participating in small rooms only or has few peers, the chance to successfully attack the scheme is high, as each share becomes more important. Therefore, it has to be considered, whether such users and their \storages~are worth protecting. The meaningfulness of the scheme is only given to a user if he is part of a certain number of chatrooms and knows enough peers.
\subsection{Advantages of Partitioned Storages}
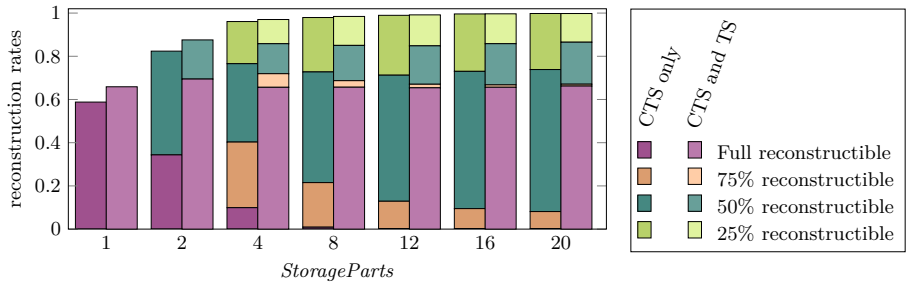
\begin{figure}[htb]
\centering
\resizebox{\textwidth}{!}{
\begin{tikzpicture}[
  every axis/.style={ % add these settings to all the axis environments in the tikzpicture
      ybar stacked,
      ymin=0,ymax=1.02,
      xmin=-.5, xmax=6.6,
      %xtick=(0,1,2,3,4,5,6),
      xticklabels={0,1,2,4,8,12,16,20},
      xtick style={draw=none},
      %x tick label style={rotate=0,anchor=east},
      bar width=14pt,
      scale only axis=true,
      width=0.7\textwidth,
      height=3.5cm,
      		ylabel=reconstruction rates,
      		xlabel=\storageParts,
      },
]

\begin{axis}[bar shift=-7pt,hide axis,      
legend columns=2,
legend style={
                legend cell align=left,
            ,at={(1.3,1.0)}, anchor=north, every even column/.append style={column sep=-30pt}}]

\addlegendimage{empty legend}
\addlegendentry{\hspace*{-5pt}\rotatebox{70}{\hspace*{-10pt}CTS only}}
\addlegendimage{empty legend}
\addlegendentry{\hspace*{-10pt}\rotatebox{70}{\hspace*{2pt}CTS and TS}}

\addlegendimage{draw=black,fill=color1a}
\addlegendentry{}
\addlegendimage{draw=black,fill=color1b}
\addlegendentry{\hspace*{5pt}Full reconstructible}
\addlegendimage{draw=black,fill=color2a}
\addlegendentry{}
\addlegendimage{draw=black,fill=color2b}
\addlegendentry{\hspace*{5pt}75\% reconstructible}
\addlegendimage{draw=black,fill=color3a}
\addlegendentry{}
\addlegendimage{draw=black,fill=color3b}
\addlegendentry{\hspace*{5pt}50\% reconstructible}
\addlegendimage{draw=black,fill=color4a}
\addlegendentry{}
\addlegendimage{draw=black,fill=color4b}
\addlegendentry{\hspace*{5pt}25\% reconstructible}

\addplot[fill=color1a, draw=black] coordinates
{(0,.58742)(1,.34298)(2,.09859)(3,.00833)(4,.00088)(5,.00018)(6,.00004)};
\addplot[fill=color2a, draw=black] coordinates
{(0,0)(1,0)(2,.30373)(3,.20626)(4,.12749)(5,.09406)(6,.08053)};
\addplot[fill=color3a, draw=black] coordinates
{(0,0)(1,.48028)(2,.3628)(3,.51234)(4,.58356)(5,.63546)(6,.65739)};
\addplot[fill=color4a, draw=black] coordinates
{(0,0)(1,0)(2,.19506)(3,.25189)(4,.27708)(5,.26569)(6,.25946)};
\end{axis}

\begin{axis}[bar shift=7pt]
\addplot+[fill=color1b, draw=black] coordinates
{(0,.65887)(1,.69459)(2,.65644)(3,.65685)(4,.65407)(5,.65665)(6,.66227)};
\addplot+[fill=color2b, draw=black] coordinates
{(0,0)(1,0)(2,.06287)(3,.03017)(4,.01649)(5,.01087)(6,.00855)};
\addplot+[fill=color3b, draw=black] coordinates
{(0,0)(1,.18118)(2,.13887)(3,.16339)(4,.1778)(5,.19076)(6,.19439)};
\addplot[fill=color4b, draw=black] coordinates
{(0,0)(1,0)(2,.11201)(3,.134)(4,.14334)(5,.13828)(6,.13271)};
\addplot[fill=color1a, draw=black] coordinates{(-5,-5)};
\addplot[fill=color2a, draw=black]  coordinates{(-5,-5)};
\addplot[fill=color3a, draw=black]  coordinates{(-5,-5)};
\addplot[fill=color4a, draw=black]  coordinates{(-5,-5)};
\end{axis}

%a 	0.58742	0	0	0
%b	0.65887	0	0	0
%a	0.34298	0	0.48028	0
%b	0.69459	0	0.18118	0
%a	0.09859	0.30373	0.3628	0.19506
%b	0.65644	0.06287	0.13887	0.11201
%a	0.00833	0.20626	0.51234	0.25189
%b	0.65685	0.03017	0.16339	0.134
%a	0.00088	0.12749	0.58356	0.27708
%b	0.65407	0.01649	0.1778	0.14334
%a	0.00018	0.09406	0.63546	0.26569
%b	0.65665	0.01087	0.19076	0.13828
%a	0.00004	0.08053	0.65739	0.25946
%b	0.66227	0.00855	0.19439	0.13271
\end{tikzpicture}
}
\caption{Adding a share per peer, for the \storage~increases the reconstruction rates, while the rate of partly reconstructible \storages~decreases.}
\label{chart:modification}
\end{figure}

In \cite{seufert2015analysis, rosenfeld2018study} two studies about the usage of WhatsApp were performed. The usage behavior from these studies was used to perform simulations about the advantages of partitioned storages. According to the studies, an average user of an online chat has contact with around 70 peers, in about 60 different chats. About 71.5\% of the chats are between 2 peers, 11.4\% are between 3 to 5 peers, 6.9\% are between 6 to 10 peers, and the remaining chats contain 11 peers or more. Peers can be in multiple chats, which explains the fact, that the number of peers and chats does not differ that much. These statistics were applied to different simulations to find advantages, disadvantages, and optimal values when implementing our approach.\\
For most simulations, a target threshold rate of $t_{\mathit{target}}=0.7$ was used. I.e. the threshold rate of the \storage~and \storageParts~satisfy the equation: \\${t_{\mathit{storage}}\cdot t_{\mathit{storagePart}} = t_{\mathit{target}}}$. This means that a low fraction of 70\% or more of the shareholders should be able to reconstruct the \storage. This fraction was chosen, because update procedures ensure, that only active peers receive shares. This leads to a low amount of inactive peers. It is far more common for users to forget their passwords when they are not using the services for a while. Therefore, a large number of peers could be inactive in the meanwhile. No research was found that allowed us to estimate a better value. To generate appropriate simulations, therefore, 70\% of the peers are marked as inactive, using a uniformly distributed random number generator. Inactive peers cannot help in reconstructing the secret. For all simulations, the reconstruction rates are calculated. A reconstruction rate of $r=0.9$ means that in 90\% of the simulations the \storage~and all \storagePart~can be reconstructed from the available shares. Further, partly reconstruction rates were calculated. I.e. a 75\%-reconstruction rate gives the rate of simulations where between 50\% and 75\% of the \storageParts~can be reconstructed. In some simulations, a reconstruction rate $r_a=r+r_{75\%}+r_{50\%}+r_{25\%}$ is used, so any instance where at least 25\% of the \storageParts~were successfully reconstructed are counted.\\
\textbf{Is it Beneficial to Consider Unique Peers?}
One of the questions is, whether it is beneficial to share the secret with unique peers in a \storageParts, versus distributing shares to peers for every occurrence in the chats in the \storagePart. I.e. should a peer $p$, who is a participant in two chatrooms $c_1$ and $c_2$, where both chatrooms are saved in the same \storagePart, receive one unique share $s_p$ or two different shares $s_{p,{c_1}}$ and $s_{p,{c_2}}$? Figure~\ref{chart:uniquepeers} shows the results of simulations for scenarios where the number of \storageParts~varies between 1 and 8. For a single \storagePart~a significant advantage, when considering unique peers, is visible. For all other numbers of \storageParts, the advantage turns out lower but is still there. Therefore, considering unique peers brings slight advantages. For one single \storagePart, the reconstruction rate $r_t$ increases from $0.59$ to $0.67$. For two \storageParts~the increase is from $0.81$ to $0.82$.
\begin{figure}[htb]
\centering
\begin{tikzpicture}
%	\begin{axis}[
%	scale only axis=true,
%	width=0.6\textwidth,
%	height=2.2cm,
%		ybar, 	
%		ylabel=reconstruction rate,
%		xlabel=\storageParts,
%		xtick={0,1,2,3,4,5,6,7},
%		xticklabels={1,2,3,4,5,6,7,8},
%		legend style={at={(1.25,1.0)},
%		anchor=north},
%		bar width=8pt,
%		xtick style={draw=none},
%		]
%	\addplot[fill=color1b, draw=black] coordinates
%		{(0,0.58956) (1,.8112) (2,.91427) (3,.96119) (4,.98198) (5,.9917) (6,.99602) (7,.99823) };
%			\addplot[fill=color1a, draw=black] coordinates
%				{(0,0.66826) (1,.82342) (2,.91529) (3,.96084) (4,.98196) (5,.99155) (6,.99602) (7,.99821) };
%	\legend{Non-unique Peers, Unique Peers}
%	\end{axis}
%%0.66826	0.58956
%%0.82342	0.8112
%%0.91529	0.91427
%%0.96084	0.96119
%%0.98196	0.98198
%%0.99155	0.9917
%%0.99602	0.99602
%%0.99821	0.99823

\begin{axis}[
xlabel=\storageParts,
ylabel=reconstruction rate,
ylabel style={xshift=-0.8em},
	scale only axis=true,
	width=.85\textwidth,
	height=2.3cm,
	legend style={
		at={(0.8,0.6)},
		anchor=north,},
		%legend columns=-1},
]
\addplot[draw=color1b,mark=square*,mark options={fill=color1b}] coordinates {
	(1,.66826)
	(2,.82342)
	(3,.91529)
	(4,.96084)
	(5,.98196)
	(6,.99155)
	(7,.99602)
	(8,.99821)
};

\addplot[draw=color1a,mark=triangle*,mark options={fill=color1a}] coordinates {
	(1,.58956)
	(2,.8112)
	(3,.91427)
	(4,.96119)
	(5,.98198)
	(6,.9917)
	(7,.99602)
	(8,.99823)
};
\legend{Non-unique Peers, Unique Peers}
\end{axis}

\end{tikzpicture}
\caption{The rate of total and partly reconstructible \storages, when distributing single shares to unique peers instead of multiple shares per peer, is higher for low numbers of \storageParts. The difference decreases with additional \storageParts.}
\label{chart:uniquepeers}
\end{figure}
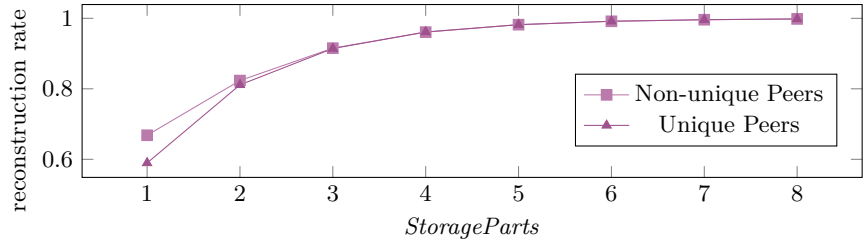\\
\textbf{Can the Recoverability be Improved with the Additional \TSSS?}
In Figure~\ref{chart:modification} the impact on the reconstruction rates when using a unique share per peer for the \storage~is shown. I.E. two different secret sharing schemes \TSSS~and \CTSSS~are used. Every first bar shows the reconstruction rates when peers receive additional shares for reconstructing the \storage~(\TSSS), the second bar shows the reconstruction rates, without this additional share (\CTSSS~only). For every number of \storageParts, a significant increase in $r$ is visible. At the same time, the combined partly reconstruction rates decrease. It is visible, that the introduction of additional shares, improves the scheme significantly. For a single \storagePart, an increase from 0.59 to 0.66 was measured. For two \storageParts, the increase for full recoverability is from 0.34 to 0.69, $r_{50\%}$ drops from 0.48 to 0.18. For four \storageParts, the full recoverability increases from 0.1 to 0.66, $r_{75\%}$ drops from 0.3 to 0.06, $r_{50\%}$ drops from 0.36 to 0.14, and $r_{25\%}$ drops from 0.2 to 0.11.
\\
\textbf{Should a $(p,p)$-Threshold Scheme be Used?}
A target threshold rate of $t_{\mathit{target}}=0.7$ allows to vary the threshold rates $t_{\mathit{storage}}$ and $t_{\mathit{storagePart}}$, as long as the product of both equals $t_{\mathit{target}}$. Therefore, to find the optimal value for both threshold rates the simulations shown in Figure~\ref{chart:tt-threshold} were performed. For three different numbers of \storageParts, it was simulated, which value $q \in \{p, p-1, p-2\}$ should optimally be used for the $(p,q)$-threshold scheme to reconstruct the \storage~from the \storageParts. The left bars show the results for $q=p$, the middle bars show the results for $q = p-1$, and the right bars show the results of $q = p-2$. Clearly visible is, that $r$ stays nearly equal with varying $q$, whereas the partly reconstruction rates drop significantly. This result seems likely, because decreasing $t_{\mathit{storage}}$ results in an increasing $t_{\mathit{storagePart}}$, which results in less reconstructed \storageParts, whereas the \storages~still can be reconstructed using the \TSSS, as seen in Figure~\ref{chart:modification}. Therefore, using a $(p,p)$-\TSSS~seems optimal.
\begin{figure}[htb]
\centering

	\resizebox{\textwidth}{!}{

		\begin{tikzpicture}[
		every axis/.style={ % add these settings to all the axis environments in the tikzpicture
			ybar stacked,
			ymin=0.5,ymax=1.02,
			xmin=-.5, xmax=10.5,
			xtick={0,1,2,3,4,5,6,7,8,9,10},
			xticklabels={$\frac{p}{p}$,$\frac{p}{p-1}$,$\frac{p}{p-2}$,,$\frac{p}{p}$,$\frac{p}{p-1}$,$\frac{p}{p-2}$,,$\frac{p}{p}$,$\frac{p}{p-1}$,$\frac{p}{p-2}$},
			xtick style={draw=none},
			%x tick label style={rotate=0,anchor=east},
			bar width=20pt,
			scale only axis=true,
			width=0.60\textwidth,
			height=3.2cm,
			ylabel=reconstruction rates,
			xlabel=total threshold,
		},
		]
		
		\begin{axis}[  
			legend columns=3,
			legend style={
				legend cell align=left,
				,at={(1.3,0.75)}, 
				anchor=north,/tikz/every even column/.append style={column sep=-5pt}}]% /immer der erste bar
			\addlegendimage{draw=black,fill=color1a}
			\addlegendentry{}
			\addlegendimage{draw=black,fill=color1b}		
			\addlegendentry{}
			\addlegendimage{draw=black,fill=color1c}
			\addlegendentry{Full reconstructible}
			\addlegendimage{draw=black,fill=color2a}
			\addlegendentry{}
			\addlegendimage{draw=black,fill=color2b}
			\addlegendentry{}
			\addlegendimage{draw=black,fill=color2c}
			\addlegendentry{75\% reconstructible}
			\addlegendimage{draw=black,fill=color3a}
			\addlegendentry{}
			\addlegendimage{draw=black,fill=color3b}
			\addlegendentry{}
			\addlegendimage{draw=black,fill=color3c}
			\addlegendentry{50\% reconstructible}
			\addlegendimage{draw=black,fill=color4a}
			\addlegendentry{}
			\addlegendimage{draw=black,fill=color4b}
			\addlegendentry{}
			\addlegendimage{draw=black,fill=color4c}
			\addlegendentry{25\% reconstructible}

		\addplot[fill=color1a, draw=black] coordinates
		{(0,.66136)(1,0)(2,0)(3,0)(4,.65271)(5,0)(6,0)(7,0)(8,.65543)(9,0)(10,0)};
		\addplot[fill=color2a, draw=black] coordinates
		{(0,0.02871)(1,0)(2,0)(3,0)(4,0.01553)(5,0)(6,0)(7,0)(8,.01114)(9,0)(10,0)};
		\addplot[fill=color3a, draw=black] coordinates
		{(0,0.16183)(1,0)(2,0)(3,0)(4,.17884)(5,0)(6,0)(7,0)(8,.19101)(9,0)(10,0)};
		\addplot[fill=color4a, draw=black] coordinates
		{(0,0.1328)(1,0)(2,0)(3,0)(4,0.14513)(5,0)(6,0)(7,0)(8,.13847)(9,0)(10,0)};
	
	%%0a 0.66136	0.02871	0.16183	0.1328
	%%0b 0.66194	0.00001	0.00519	0.07887
	%%0c 0.657		0		0.00234	0.02085
	%			
	%%1a 0.65271	0.01553	0.17884	0.14513
	%%1b 0.65682	0.00053	0.03427	0.20433
	%%1c 0.66203	0.00013	0.00847	0.06296
	%			
	%%2a 0.65543	0.01114	0.19101	0.13847
	%%2b 0.66551	0.00269	0.09401	0.20924
	%%2c 0.65027	0.00122	0.02871	0.13192
		
		\addplot[fill=color1b, draw=black] coordinates
		{(0,0)(1,0.66194)(2,0)(3,0)(4,0)(5,0.65682)(6,0)(7,0)(8,0)(9,0.66551)(10,0)};
		\addplot[fill=color2b, draw=black] coordinates
		{(0,0)(1,0.00001)(2,0)(3,0)(4,0)(5,0.00053)(6,0)(7,0)(8,0)(9,0.00269)(10,0)};
		\addplot[fill=color3b, draw=black] coordinates
		{(0,0)(1,0.00519)(2,0)(3,0)(4,0)(5,0.03427)(6,0)(7,0)(8,0)(9,0.09401)(10,0)};
		\addplot[fill=color4b, draw=black] coordinates
		{(0,0)(1,0.07887)(2,0)(3,0)(4,0)(5,0.20433)(6,0)(7,0)(8,0)(9,0.20924)(10,0)};
		
		\addplot[fill=color1c, draw=black] coordinates
		{(0,0)(1,0)(2,0.657)(3,0)(4,0)(5,0)(6,0.66203)(7,0)(8,0)(9,0)(10,0.65027)};
		\addplot[fill=color2c, draw=black] coordinates
		{(0,0)(1,0)(2,0)(3,0)(4,0)(5,0)(6,0.00013)(7,0)(8,0)(9,0)(10,0.00122)};
		\addplot[fill=color3c, draw=black] coordinates
		{(0,0)(1,0)(2,0.00234)(3,0)(4,0)(5,0)(6,0.00847)(7,0)(8,0)(9,0)(10,0.02871)};
		\addplot[fill=color4c, draw=black] coordinates
		{(0,0)(1,0)(2,0.02085)(3,0)(4,0)(5,0)(6,0.06296)(7,0)(8,0)(9,0)(10,0.13192)};

		\end{axis}
		\node[] at (1,3.5) {8~\storageParts};
		\node[] at (3.67,3.5) {12~\storageParts};
		\node[] at (6.35,3.5) {16~\storageParts};
		\end{tikzpicture}
	}

\caption{The rates for full recovery stays nearly the same when decreasing the threshold for the reconstruction of the \storage. The rates for 75\% and 50\% recoveries decrease in most combinations.}
\label{chart:tt-threshold}
\end{figure}
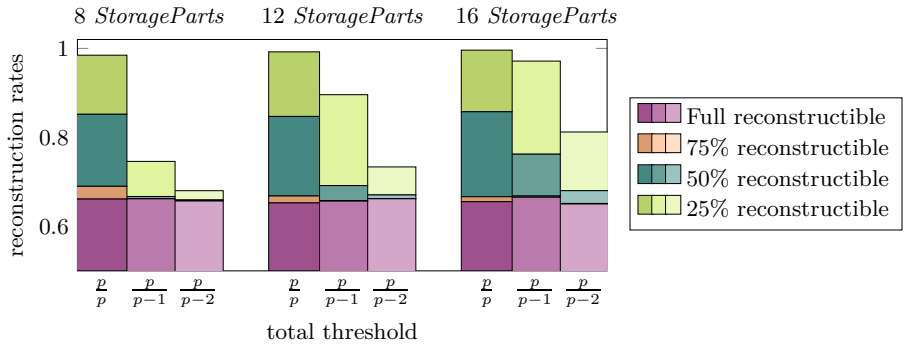\\
\textbf{What is a Good Threshold Value?} In Figure~\ref{chart:90threshold} the results of simulations with target threshold values $t_{\mathit{target}}=0.9$ and $t_{\mathit{target}}=0.7$ and different numbers of \storageParts~are shown. The overall behavior of the scheme stays the same for different target threshold values, with slightly higher reconstruction rates for $t_{\mathit{target}}=0.9$. The reconstruction rates $r$, for $t_{\mathit{target}}=0.9$ are 0.73 for a single \storagePart, 0.75 for two \storageParts, and 0.72 for four \storageParts. $r_{50\%}$ is at 0.16 for two \storageParts, and at 0.12 for four \storageParts. For four \storageParts, $r_{75\%}$ is 0.06. For $t_{\mathit{target}}=0.7$ the values are slightly lower.
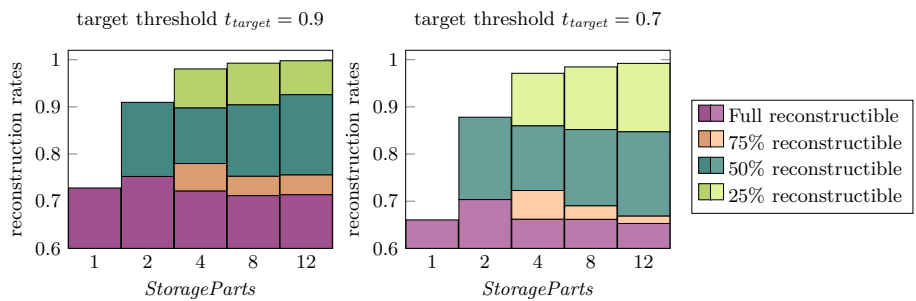
\begin{figure}
	
	\centering
	\resizebox{\textwidth}{!}{

			\begin{tikzpicture}[
			every axis/.style={ % add these settings to all the axis environments in the tikzpicture
				ybar stacked,
				ymin=0.6,ymax=1.02,
				xmin=-.5, xmax=4.5,
				xtick={0,1,2,3,4},
				xticklabels={1,2,4,8,12},
				xtick style={draw=none},
				%x tick label style={rotate=0,anchor=east},
				bar width=24pt,
				scale only axis=true,
				width=0.35\textwidth,
				height=3.2cm,
				ylabel=reconstruction rates,
				xlabel=\storageParts,
			},
			]
			
			\begin{axis}[  
			title={target threshold $t_{\mathit{target}}=0.9$}   
			] 
			%	legend columns=1,
			%	legend style={
			%		legend cell align=left,
			%		,at={(0.5,-.32)}, anchor=north,/tikz/every even column/.append style={column sep=0.5cm}}] /immer der erste bar
			%	\addlegendimage{empty legend}
			%	\addlegendentry{\hspace*{-8pt}target threshold $t_{\mathit{target}}=0.9$};
			%	
			%	\addlegendimage{draw=black,fill=color1a}
			%	\addlegendentry{Full reconstructible}
			%	\addlegendimage{draw=black,fill=color2a}
			%	\addlegendentry{75\% reconstructible}
			%	\addlegendimage{draw=black,fill=color3a}
			%	\addlegendentry{50\% reconstructible}
			%	\addlegendimage{draw=black,fill=color4a}
			%	\addlegendentry{25\% reconstructible}
			
			\addplot[fill=color1a, draw=black] coordinates
			{(0,.72802)(1,.75239)(2,.72131)(3,.71176)(4,.71357)};
			\addplot[fill=color2a, draw=black] coordinates
			{(0,0)(1,0)(2,.05852)(3,.04095)(4,.04198)};
			\addplot[fill=color3a, draw=black] coordinates
			{(0,0)(1,.15723)(2,.11774)(3,.15157)(4,.1702)};
			\addplot[fill=color4a, draw=black] coordinates
			{(0,0)(1,0)(2,.0829)(3,.08841)(4,.07208)};
			
			\end{axis}

			%0a	0.72802	0		0		0
			%0b	0.6603	0		0		0
			%1a	0.75239	0		0.15723	0
			%1b	0.70337	0		0.1747	0
			%2a	0.72131	0.05852	0.11774	0.0829
			%2b	0.66156	0.06107	0.13707	0.11151
			%3a	0.71176	0.04095	0.15157	0.08841
			%3b	0.66136	0.02871	0.16183	0.1328
			%4a	0.71357	0.04198	0.1702	0.07208
			%4b	0.65271	0.01553	0.17884	0.14513
			\end{tikzpicture}
	\begin{tikzpicture}[
			every axis/.style={ % add these settings to all the axis environments in the tikzpicture
				ybar stacked,
				ymin=0.6,ymax=1.02,
				xmin=-.5, xmax=4.5,
				xtick={0,1,2,3,4},
				xticklabels={1,2,4,8,12},
				xtick style={draw=none},
				%x tick label style={rotate=0,anchor=east},
				bar width=24pt,
				scale only axis=true,
				width=0.35\textwidth,
				height=3.2cm,
				ylabel=reconstruction rates,
				xlabel=\storageParts,
			},
			]
			
			\begin{axis}[
			title={target threshold $t_{\mathit{target}}=0.7$},    
			legend columns=2,
			legend style={
				legend cell align=left,
				,at={(1.5,0.75)}, anchor=north,/tikz/every even column/.append style={column sep=-5pt}}]% /immer der erste bar
			%\addlegendimage{empty legend}
			%\addlegendentry{\hspace*{-8pt}target threshold $t_{\mathit{target}}=0.7$}
			%
			%\addlegendimage{draw=black,fill=color1b}
			%\addlegendentry{Full reconstructible}
			%\addlegendimage{draw=black,fill=color2b}
			%\addlegendentry{75\% reconstructible}
			%\addlegendimage{draw=black,fill=color3b}
			%\addlegendentry{50\% reconstructible}
			%\addlegendimage{draw=black,fill=color4b}
			%\addlegendentry{25\% reconstructible}
			
			%	legend columns=2,
			%	legend style={
			%	legend cell align=left,
			%	,at={(0.5,-.32)}, anchor=north,/tikz/every even column/.append style={column sep=0.5cm}}]
			%\addlegendimage{empty legend}
			%\addlegendentry{\hspace*{-8pt}target threshold $t_{\mathit{target}}=0.9$}
			%\addlegendimage{empty legend}
			%\addlegendentry{\hspace*{-8pt}target threshold $t_{\mathit{target}}=0.7$}
			%
			\addlegendimage{draw=black,fill=color1a}
			\addlegendentry{}
			\addlegendimage{draw=black,fill=color1b}
			\addlegendentry{Full reconstructible}
			\addlegendimage{draw=black,fill=color2a}
			\addlegendentry{}
			\addlegendimage{draw=black,fill=color2b}
			\addlegendentry{75\% reconstructible}
			\addlegendimage{draw=black,fill=color3a}
			\addlegendentry{}
			\addlegendimage{draw=black,fill=color3b}
			\addlegendentry{50\% reconstructible}
			\addlegendimage{draw=black,fill=color4a}
			\addlegendentry{}
			\addlegendimage{draw=black,fill=color4b}
			\addlegendentry{25\% reconstructible}
			
			\addplot[fill=color1b, draw=black] coordinates
			{(0,.6603)(1,.70337)(2,.66156)(3,.66136)(4,.65271)};
			\addplot[fill=color2b, draw=black] coordinates
			{(0,0)(1,0)(2,.06107)(3,.02871)(4,.01553)};
			\addplot[fill=color3b, draw=black] coordinates
			{(0,0)(1,.1747)(2,.13707)(3,.16183)(4,.17884)};
			\addplot[fill=color4b, draw=black] coordinates
			{(0,0)(1,0)(2,.11151)(3,.1328)(4,.14513)};
			\end{axis}

			%0a	0.72802	0		0		0
			%0b	0.6603	0		0		0
			%1a	0.75239	0		0.15723	0
			%1b	0.70337	0		0.1747	0
			%2a	0.72131	0.05852	0.11774	0.0829
			%2b	0.66156	0.06107	0.13707	0.11151
			%3a	0.71176	0.04095	0.15157	0.08841
			%3b	0.66136	0.02871	0.16183	0.1328
			%4a	0.71357	0.04198	0.1702	0.07208
			%4b	0.65271	0.01553	0.17884	0.14513
			\end{tikzpicture}}

	%stacked	
	%0a	0.72802	0.72802	0.72802	0.72802
	%0b	0.6603	0.6603	0.6603	0.6603
	%1a	0.75239	0.75239	0.75239	0.90962
	%1b	0.70337	0.70337	0.70337	0.87807
	%2a	0.72131	0.77983	0.89757	0.98047
	%2b	0.66156	0.72263	0.8597	0.97121
	%3a	0.71176	0.75271	0.90428	0.99269
	%3b	0.66136	0.69007	0.8519	0.9847
	%4a	0.71357	0.75555	0.92575	0.99783
	%4b	0.65271	0.66824	0.84708	0.99221
	
	\caption{Comparisons of different amounts of \storageParts~$p$ show that good values are $p=2$ and $p=4$. Especially when considering lower values $t_{\mathit{target}}$ the amount of full or 75\% partial reconstructible~\storages~decreases with larger $p$.}
	\label{chart:90threshold}
\end{figure}\\
\textbf{How many Partitions are Optimal?}
Either two or four \storageParts~are optimal, according to the simulation results shown in Figure~\ref{chart:modification}. When using two \storageParts~the rate $r$ is slightly higher, compared to four \storageParts. But, taking in to account the partial reconstruction rates especially for 75\% partial reconstruction, four \storageParts~seem to perform better. When looking at the higher target threshold of $t_{\mathit{target}}=0.9$, as shown in Figure~\ref{chart:90threshold}, the difference between both total reconstruction rates is even less. Therefore, if the overhead in disk space and communication messages can be neglected four \storageParts~are the best choice.

\section{Conclusion}
\label{Conclusion}
We have introduced a novel approach to partition encrypted private user storages, which are used in end-to-end encrypted OSNs. The partitions consist of collections of chatrooms. Each peer in each chatroom receives multiple shares for a compartmented secret sharing scheme and an additional threshold secret sharing scheme. The shares can be used to reconstruct the partitions of the user storage, which then can be used to reconstruct the complete user storage. The user storage can be recovered through a \TSSS, as well, using another set of shares. We performed simulations to find the optimal threshold values and the optimal number of partitions. Using these values, a high probability to reconstruct either the complete user storage or as many parts and chats of it is achieved. This is especially interesting when high numbers of shares are inaccessible because participants become inactive or leave the OSN. Further, different attack scenarios on the scheme were analyzed.  
\subsection*{Acknowledgments}
The authors acknowledge the financial support by the Federal Ministry of Education and Research of Germany in the framework of SoNaTe (project number 16SV7405).
%anonymized
\bibliography{partitionedstorage}
\bibliographystyle{splncs04}
\end{document}